\documentstyle[prcrc,11pt,fleqn]{article}

\newcommand{\be}{\begin{equation}} 
\newcommand{\ee}{\end{equation}} 
\newcommand{\lb}{\left[} 
\newcommand{\rb}{\right]}
 
\newcommand{\alt}{\stackrel{<}{\sim}}

\title{\vskip -5cm \begin{flushleft} {\bf\normalsize  BROWN-HET-1119} 
\end{flushleft} \vskip 2cm 
Astrophysical constraints on primordial black hole 
formation from collapsing cosmic strings\thanks{To appear in the 
Proceedings of the Third UCLA Dark Matter Symposium and Primordial 
Black Hole Workshop 
(Marina del Rey, CA, February 1998)}}

\author{Ubi F. Wichoski\address{Department of Physics, 
        Brown University, \\ Providence, RI 
        02912, USA}
        Jane H. MacGibbon\address{Code SN3, NASA Johnson Space 
               Center, \\ Houston, TX 77058, USA} and 
        Robert H. Brandenberger$^{\mbox{\scriptsize a}}$}

\begin{document}
\maketitle

\begin{abstract}
Primordial Black Holes (PBH) may have formed from the collapse of 
cosmic string loops. The spectral shape of the PBH mass spectrum 
can be determined by the scaling argument for string networks. 
Limits on the spectral amplitude derived from extragalactic 
$\gamma$-ray and galactic $\gamma$-ray and cosmic ray flux 
observations as well 
as constraints from the possible formation of stable black 
holes remnants are reanalyzed. The 
new constraints are remarkably close to those derived from the 
normalization of the cosmic string model to the cosmic microwave 
background anisotropies. 
\end{abstract}

\section{Introduction}

Cosmic strings (CS) are linear topological defects that 
are believed to originate during 
phase transitions in the very early Universe 
\cite{VSrev,HKrev,RBrev}. 
Here, we consider the 
``standard" CS model \cite{Zel,V81}, according to 
which the network of linear defects quickly 
reaches a ``scaling" solution characterized by having the 
statistical properties of the string distribution independent of 
time if all lengths are scaled to the Hubble radius 
($R_H = c t$, where $c$ is the speed of light). 
Cosmic string loops (CSL) are continually formed by 
the intersection and self-intersection of long CS 
(infinite CS or CSL with radius of curvature 
larger than $R_H$). After formation, a loop 
oscillates due its own tension and slowly decays 
by emitting gravitational radiation. The 
initial length of a CSL is $l(t) = \alpha R_H$, 
where $\alpha$ is expected to be $\sim G \mu/c^2$. 
The mass of a CSL is 
$m(t) = l(t) \mu$, where $\mu$ 
is the mass per unit of length of the string 

Since CS also lead to cosmic microwave background 
(CMB) anisotropies, the string model can be 
normalized by the recent COBE observations 
giving the constraint \cite{LP93,ACSSV} 
\be \label{cmbnorm} 
G\mu / c^2 \leq 1.7(\pm0.7) \times 10^{-6} 
\ee
 
Our assumption is 
that a distribution of PBH was formed by the collapse of a 
fraction $f$ of CSL \cite{SH,AP}. 
Hence, from the observational consequences of a present 
surviving distribution of PBH we can derive updated 
constraints on the CS scenario \cite{jru}. 
These constraints are important because: 
{\it i-)} They may indicate new ways to search for direct 
signatures from CS; 
{\it ii-)} They may provide constraints on CS models with 
symmetry breaking scale $\mu^{1/2}$ smaller than 
$10^{16}$ GeV which are not constrained by CMB and 
large-scale structure data; and 
{\it iii-)} They may provide tighter limits than the CMB on 
CS models with $G \mu / c^2 \sim 10^{-6}$. 
 
Because CS do not dominate the energy density of the Universe, 
the CS network must lose energy. 
We derive the rate of CSL production ${{dn_l} \over {dt}}$ 
from the conservation of 
string energy in the ``scaling" scenario, 
\[  \label{consmass}
{\dot \rho_{\infty}} - 2 H \rho_{\infty} = - {{dn_l} \over {dt}} 
\alpha \mu t \, ,
\] 
where $\rho_{\infty} = \nu \mu c^{-3} t^{-2}$ is the 
energy density in long strings and $\nu$ is proportional to the 
average number of long strings crossing each Hubble 
volume. 

Hawking \cite{SH} and Polnarev and Zembowicz \cite{AP} 
first postulated that a fraction $f$ of the CSL could collapse 
within its Schwarzchild radius and then form a BH.
More recently, Caldwell and Casper \cite{CC} have performed numerical 
simulations to determine $f$ and found 
\be \label{fvalue}
f = 10^{4.9 \pm 0.2} (G \mu / c^2)^{4.1 \pm 0.1} \, .
\ee 

The BH are sufficiently small that they lose mass due to the Hawking 
evaporation process. The fraction of the critical 
density of the Universe in PBH today due to collapsing 
CSL is (see \cite{jru} and references quoted therein) 
\be \label{omegaPBH}
\Omega_{PBH}(t_o) = \frac{1}{\rho_{crit}(t_o)} 
\int_{t_*}^{t_o} dt' \frac{dn_{BH}}{dt'} m(t',t_o) \, ,
\ee 
where $t_o$ is the present age of the Universe; 
$t_*$ is formation time for a PBH with mass  
$M_* = 4.4 \times 10^{14} h^{-0.3} \; \mbox{g}$, 
which is expiring today; 
$m(t',t_0)$ is 
the mass of a PBH formed at a time $t'$ at a later time $t$; and 
$h$ is the Hubble parameter in units of 
$100 \mbox{km} \mbox{s}^{-1} \mbox{Mpc}^{-1}$. 

PBH formed at times 
$t < t_* \; (M < M_*)$ do not contribute to this integral because 
they will have evaporated by today. If we 
assume for simplicity that PBH with 
mass $M > M_*$ will have evaporated little by the present time, 
we can approximate $m(t',t_0)$ by $\alpha \mu c t'$. 

\section{Galactic and extragalactic flux constraints}

It is well known \cite{car75,PH,car76} that the extragalactic 
$\gamma$-ray flux observed at $100MeV$ provides a strong constraint 
on the population of 
black holes evaporating today. Too many black holes would lead to an 
excess of such radiation above the observed value. 
In particular, it was shown that if the present day number density 
distribution of black holes of mass $M$ is proportional to 
\cite{car75} $M^{-2.5}$ for PBH formed in the radiation-dominated 
era, then comparing the Hawking emission from the black 
hole distribution with the $\gamma$-ray background observed by 
the EGRET experiment 
implies the limit on the present black hole density 
of \cite{jane98} 
\be 
\Omega_{PBH} \alt (5.1 \pm 1.3) \times 10^{-9} h^{-1.95 \pm 0.15} \, . 
\label{newomegalimit}
\ee 

From the rate of formation of CSL it follows 
\be  
{{dn_{BH}} \over {dt}} = f {{dn_l} \over {dt}} = 
\frac{\nu}{\alpha} f c^{-3} t_{eq}^{-1/2} t^{-5/2} 
\label{formationrate}
\ee
which is also proportional to $M^{-2.5}$. Thus, we can apply the limit 
(\ref{newomegalimit}) to (\ref{omegaPBH}). Taking into 
account the formation rate of CSL given by 
(\ref{formationrate}) we can 
determine an upper bound on 
the fraction $f$ of CSL which collapses to form PBH 
\cite{jru} 
\begin{eqnarray} 
f & \alt & 6.8 (\pm 1.7) \times 10^{-19} \lb\frac{\nu}{40}\rb^{-1} 
\lb\frac{\gamma}{100}\rb^{-\frac12} 
\lb\frac{M_*}{4.4 \times 10^{14} \, h^{-0.3} \hspace{0.2cm} 
\mbox{gm}}\rb^\frac12 
\nonumber \\ & & \times \lb\frac{G \mu / c^2}{1.7 
\times 10^{-6}}\rb^{-2} h^{-0.1 \pm 0.15} 
\lb {{t_{eq}} \over {3.2 \times 10^{10} \, h^{-4} \hspace{0.2cm} 
\mbox{sec} }} \rb^{-1/2} \, , \label{fconstr2} 
\end{eqnarray} 
where $\gamma$ is a 
dimensionless coefficient describing the strength of gravitational 
radiation generated by string loops 
($\alpha = \gamma G \mu / c^2$). 

Now, if we assume the validity of the Caldwell and 
Casper simulations \cite{CC}, we can deduce an upper bound on the 
value of $G \mu / c^2$. Combining (\ref{fvalue}) and 
(\ref{fconstr2}) and taking into account (\ref{newomegalimit}), 
we have 
\be \label{gmuconstr}
G \mu / c^2 \, \leq \, 2.1(\pm 0.7) \times 10^{-6} \, .
\ee
This limit is very close to those from the normalization of the 
CS model to the CMB.

We can also apply the limits on 
$\Omega_{PBH}(t_o)$ 
coming from the observations of the Galactic $\gamma$-ray, 
antiproton, electron and 
positron fluxes. These limits, however, are less certain than 
the diffuse extragalactic $\gamma$-ray flux because of the 
dependence on the unknown degree to which PBH cluster in the Galaxy 
and on the propagation and modulation of emitted particles in the 
Galaxy and Solar System. 
Assuming a 
halo model in which the spatial distribution of black holes is 
proportional to the isothermal density distribution of dark 
matter within the Galactic halo and simulating the diffusive 
propagation of antiprotons in the Galaxy, 
Maki et al. \cite{mak} derive an upper limit on 
$\Omega_{PBH}$ of 
\[ \label{crlimit}
\Omega_{PBH} \, < \, 1.8 \times 10^{-9} h^{-4/3}
\]                
based on antiproton data from the BESS '93 balloon 
flight. This value would imply a 
limit on  $f$ in (\ref{fconstr2}) 
that is stronger by a factor of about 
3 and a corresponding limit on 
$G\mu / c^2$ in (\ref{gmuconstr}) of 
\[ \label{gmuconstr2}
G\mu / c^2 \, <  \, 1.8(\pm0.5) \times 10^{-6} \, .   
\] 

\section{Limits on black hole remnants} 

The final stage of an expiring BH is unknown. It is possible that 
the evaporation may stabilize at or before the BH mass reaches the 
Planck mass, $m_{pl}$. Thus, a mass $M_{relic}$ would remain, 
implying that the present fraction of the critical 
density in BH relics is 
\[ \label{relic2}
\Omega_{relic} \, = \, {{M_{relic}} \over {m_{pl}}} {{m_{pl}} \over 
{\rho_{crit}(t_0)}}\int_{t_i}^{t'(t_0)} dt'{{dn_{BH}} 
\over {dt'}} \, . 
\]  
For $\Omega_{relic} \leq 1$, we obtain the constraint 
\begin{eqnarray} \label{bound3} 
f \, &\leq& \, 1.9 \times 10^{-15} \bigl( {\nu \over {40}} \bigr)^{-1} 
\bigl( {{\gamma} \over {100}} \bigr)^{-1/2} \bigl( {{m_{pl}} \over 
{M_{relic}}} \bigr) \bigl( {{G \mu / c^2} \over {1.7 \times 
10^{-6}}} \bigr)^{-11/4} \; \nonumber \\ & & 
\times h^2 \lb {{t_{eq}} \over {3.2 \times 
10^{10} \, h^{-4} \hspace{0.2cm} \mbox{sec} }} \rb^{-1/2} \, .
\nonumber 
\end{eqnarray}
If we make use of both $f$ and $G \mu / c^2$ from 
(\ref{fconstr2}) and (\ref{cmbnorm}) we can derive an upper 
limit on $\Omega_{relic}$ 
\[
\Omega_{relic} \, \leq \, 3.6 (\pm 0.9) \times 
10^{-4} \bigl( {{M_{relic}} 
\over {m_{pl}}} \bigr) \bigl( 
{{M_*} \over {4.4 \times 10^{14} \, h^{-0.3} \hspace{0.2cm} 
\mbox{gm}}}\bigr)^{1/2} \bigl( {{G\mu / c^2} \over {1.7 \times 
10^{-6}}}\bigr)^{3/4} \; h^{-2.1 \pm 0.15} \, .
\] 
Thus, the bound on the BH formation efficiency factor $f$ 
given by (\ref{fconstr2}) implies that BH remnants from 
collapsing CSL would contribute significantly to the dark 
matter of the Universe in the CS scenario of structure 
formation ($G\mu / c^2 \simeq 1.7 \times 10^{-6}$) 
if the BH remnants have a relic mass larger than about 
$10^3 m_{pl}$. 

\section{Conclusions}

We have taken advantage of the recent numerical 
simulations to better understand 
PBH formation. The 
observational consequences of a PBH distribution 
were used to constrain the CS scenario. 
We have found that the limits on $G \mu/c^{2}$ are comparable to 
those stemming from other criteria. 
Unless the mass of the BH remnants is larger than 
$10^3 m_{pl}$, these 
remnants will contribute negligibly to the dark matter of the 
Universe, even if the BH formation rate has the maximal 
value allowed by the $\gamma$-ray flux constraints. A remnant mass 
of $10^3 m_{pl}$, however, can arise naturally in some 
models \cite{CPW} of BH evaporation. In this case, 
cosmic strings could consistently provide an explanation for the 
origin of cosmological structure, for the dark matter, and for the 
origin of the extragalactic $\gamma$-ray  and Galactic cosmic ray 
backgrounds around $100 MeV$.


\begin{thebibliography}{9}

\bibitem{VSrev} A. Vilenkin and E.P.S. Shellard, `Strings and 
Other Topological Defects' (Cambridge Univ. Press, Cambridge, 1994).
\bibitem{HKrev} M. Hindmarsh and T.W.B. Kibble, 
{\it Rept. Prog. Phys.} {\bf 58}, 477 (1995).
\bibitem{RBrev} R. Brandenberger, {\it Int. J. Mod. Phys.} 
{\bf A9}, 2117 (1994).
\bibitem{Zel} Ya.B. Zel'dovich, {\it Mon. Not. R. astron. Soc.} 
{\bf 192}, 663 (1980).
\bibitem{V81} A. Vilenkin, {\it Phys. Rev. Lett.} {\bf 46}, 
1169 (1981). 
\bibitem{LP93} L. Perivolaropoulos, {\it Phys. Lett.} {\bf B298}, 
305 (1993).
\bibitem{ACSSV} B. Allen et al., {\it Phys. Rev. Lett.} 
{\bf 77}, 3061 (1996).
\bibitem{SH} S. Hawking, {\it Phys. Lett.} {\bf B231}, 237 (1989).
\bibitem{AP} A. Polnarev and R. Zembowicz, {\it Phys. Rev.} 
{\bf D43}, 1106 (1991).
\bibitem{jru} J. H. MacGibbon, R. Brandenberger and 
U.F. Wichoski, {\it Phys. Rev.} {\bf D57}, 2158 (1998). 
\bibitem{CC} R. Caldwell and P. Casper, {\it Phys. Rev.} {\bf D53}, 
3002 (1996).
\bibitem{car75} B.J. Carr, {\it Ap. J.} {\bf 201}, 1 (1975).
\bibitem{PH} D. Page and S. Hawking, {\it Ap. J.} {\bf 206}, 
1 (1976). 
\bibitem{car76} B.J. Carr, {\it Ap. J.} {\bf 206}, 8 (1976). 
\bibitem{jane98} J.H. MacGibbon and B.J. Carr, in preparation (1998).
\bibitem{mak} K. Maki et al., {\it Phys. Rev. Lett.} {\bf 76}, 
3474 (1996)
\bibitem{CPW} S. Coleman, J. Preskill and F. Wilczek, 
{\it Mod. Phys. Lett.} {\bf A6}, 1631 (1991).

\end{thebibliography}
\end{document}